\documentclass[preprint,prl,unsortedaddress,superscriptaddress]{revtex4}
\usepackage{amsmath,color}
\usepackage[pdftex]{graphicx}
\allowdisplaybreaks
\begin{document}
\title{Acoustic Metamaterial with Negative Modulus}

\author{Sam Hyeon Lee}
\affiliation{Institute of Physics and Applied Physics, Yonsei
University, Seoul 120-749, Korea}
\author{Choon Mahn Park}
\affiliation{AEE Center, Anyang University, Anyang 430-714, Korea}
\author{Yong Mun Seo}
\affiliation{Department of Physics, Myongji University, Yongin
449-728, Korea}
\author{Zhi Guo Wang}
\affiliation{Department of Physics, Tongji University, Shanghai
200092, People's Republic of China}

\author{Chul Koo Kim\footnote{To whom correspondence should be addressed. E-mail:
ckkim@yonsei.ac.kr}}
\affiliation{Institute of Physics and Applied
Physics, Yonsei University, Seoul 120-749, Korea}

\date\today

\begin{abstract}

\textbf{Abstract.} We present experimental and theoretical results
on an acoustic metamaterial that exhibits negative effective modulus
in a frequency range from 0 to 450 Hz. One-dimensional acoustic
metamaterial with an array of side holes on a tube was fabricated.
We observed that acoustic waves above 450 Hz propagated well in this
structure, but no sound below 450 Hz passed through. The frequency
characteristics of the metamaterial has the same form as that of the
permittivity in metals due to the plasma oscillation. We also
provide a theory to explain the experimental results.

\end{abstract}

\maketitle

Many phenomena previously regarded impossible have been realized
using metamaterials consisting of sub-wavelength unit cells.
Metamaterials with negative values of constitutive
parameters~\cite{1,2,3,4,5,6,7,8,9,10} received much attentions
because of their novelty and applicability for fascinating practices
such as optical superlensing and cloaking~\cite{11,12,13,14,15}.
Most of the negative metamaterials are based on local resonators:
Split ring resonators for $\mu$-negativity~\cite{4,5}, Helmholtz
resonators for negative modulus~\cite{6}, and a membrane resonator
for a negative dynamic mass~\cite{16}. These negative parameters
stem from the reaction of the oscillating resonant
elements~\cite{17}. Therefore the frequency ranges for the negative
parameters are limited to bands that do not extend to zero
frequency. We will refer to these negative materials as
local-resonator-type metamaterials to distinguish from the
non-local-resonator-type metamaterials. For example, the metal wires
used to generate low frequency $\epsilon$-negative material are not
resonators themselves~\cite{3,5}. Because the negative permittivity
is not due to the resonating unit elements, the frequency range for
negativity extends from a cut-off down to zero frequency. Another
example of non-local-resonator-type metamaterials is the
zero-frequency metamaterial based on superconducting blocks reported
by Wood and Pendry~\cite{18}. In this paper we present acoustic
example of the non-local-resonator-type metamaterial. This
metamaterial exhibits negative effective modulus in the frequency
range from zero to a cut-off frequency, with the frequency
characteristics of the same from as that of the metallic
permittivity.

It is well established that plasmon in metals or in an arrays of
metal wires produces dielectric frequency characteristics given by

\begin{equation} \label{eq:sound-1}
\epsilon(\omega) = \epsilon_{o} \left( 1 -
\frac{{\omega}^2_p}{\omega ( \omega + i \gamma )} \right),
\end{equation}
where ${\omega}_p$ is the plasma frequency, and the parameter
$\gamma$ is a damping term~\cite{2,3}. The phase velocity of
electromagnetic waves becomes
\begin{equation} \label{eq:sound-2}
v_{ph} = \sqrt{\frac{1}{\epsilon \mu}} = \sqrt{\frac{1}{\epsilon_o
\mu_o \left(1 - {\omega}_p^2/\omega(\omega + i \gamma) \right)}} =
\frac{v_o}{\sqrt{1 - \omega_p^2/\omega(\omega + i \gamma)}},
\end{equation}
where $v_o$ is the speed of light in vacuum.

If we neglect the damping term $\gamma$, equation (\ref{eq:sound-2})
becomes $v_{ph} = c/\sqrt{1 - \omega_p^2 /\omega^2}$: Below the
cut-off frequency $\omega_p$, the electromagnetic waves do not
propagate but decays exponentially with distance. At the plasma
frequency, the wavelength becomes infinite and all the electrons
oscillate in phase creating plasma oscillation. Above $\omega_p$,
the phase velocity decreases to approach the speed of light in
vacuum.

\begin{figure}
\begin{center}
\includegraphics*[width=0.8\columnwidth]{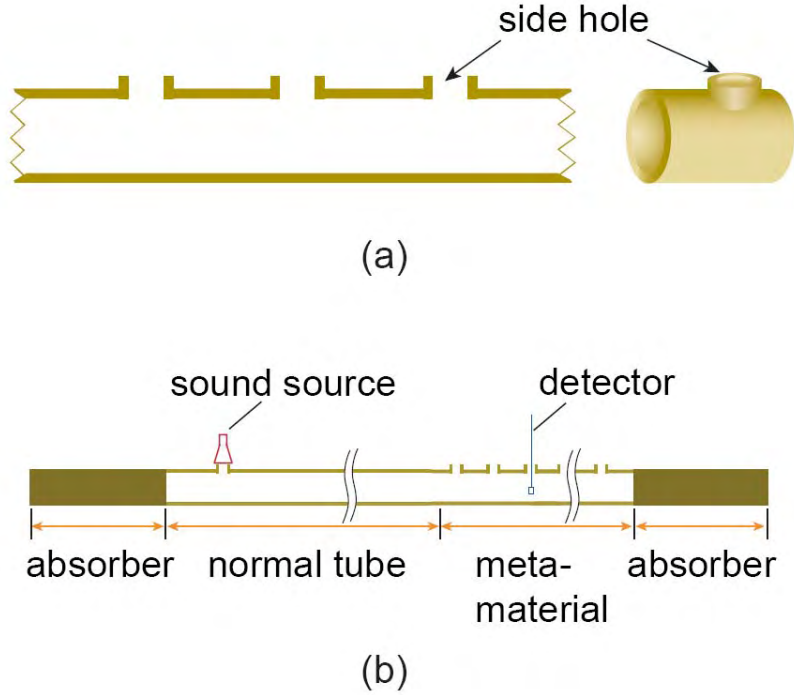}
\end{center}
\caption{(color online) (a) Structures of metamaterials;
one-dimensional structure consisting of an array of the side holes
on a tube. This system exhibits negative effective modulus. The unit
cell is shown on the right. (b) Experimental setup for the
transmission and phase velocity measurements.} \label{fig:diag}
\end{figure}

In this paper we present a homogenized acoustic metamaterial with a
negative modulus that behaves analogously to the negative
permittivity. We constructed the metamaterial as schematically shown
in figure \ref{fig:diag}(a). The unit cell is a short tube with a
side hole(SH). Unlike the Helmholz resonator, this unit cell does
not resonate acoustically by itself. When the unit cells are
connected, it becomes a tube with a regular array of SHs. The tube
has 32.3 mm inner diameter, and the SHs (diameter 10 mm) are spaced
by $d$ = 70 mm.

The experimental setup in figure \ref{fig:diag}(b) consists of a
normal tube on the left and the metamaterial on the right. The
length of the metamaterial is 1.7 m, consisting of 25 unit cells.
The absorbers at both ends completely absorb the acoustic energies
not to allow any reflections, so that the system behaves as if it
extends to infinity. The absorbers are long tubes with arrays of
sponge-like resistant plates placed inside to make the acoustic
waves decay completely. This eliminates concern about the effect of
the finite number of cells used in the experiment as well as the
interference effect from the reflected waves. The sound source
injects acoustic energy into the tube through a small hole,
generating incident waves propagating to the right. At the boundary
to the metamaterial, portion of the incident energy is reflected and
the rest transmitted. In the metamaterial side, the transmitted
acoustic energy flows steadily to the right until it hits the
absorber.

\begin{figure}
\begin{center}
\includegraphics*[width=0.45\columnwidth]{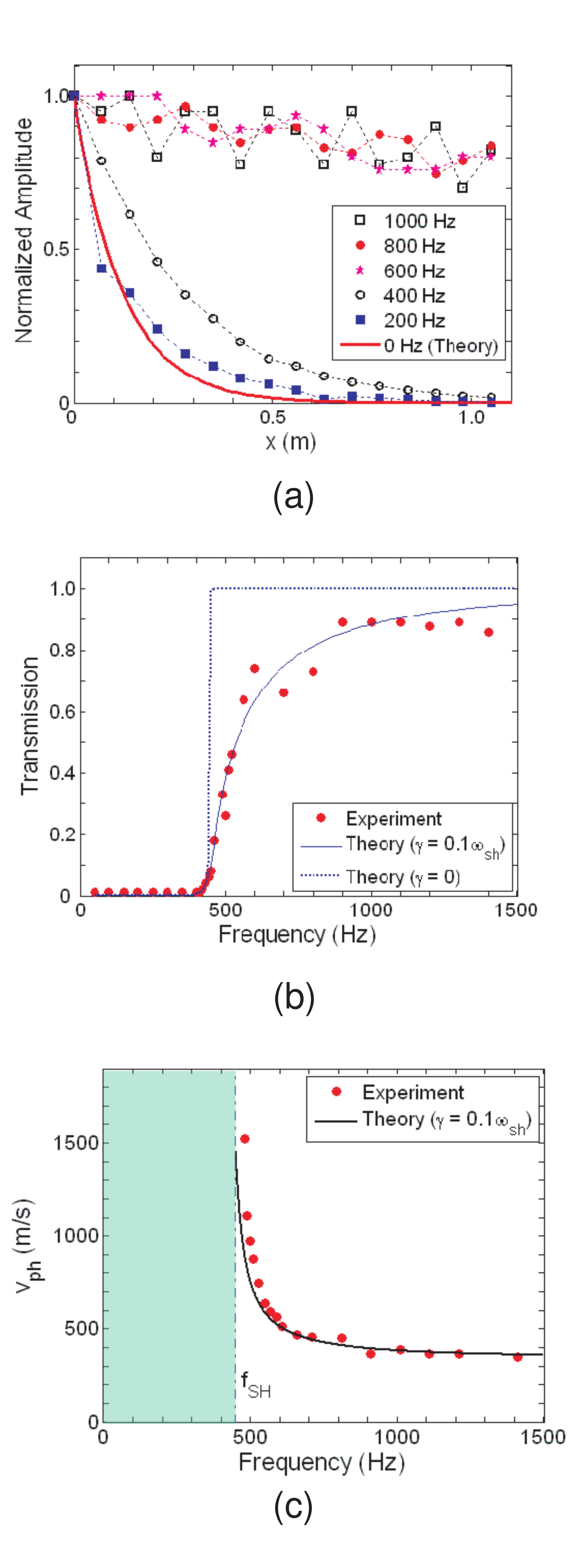}
\end{center}
\caption{(color online) (a) Sound intensities as functions of the
distance from the boundary, $x$. The broken lines connecting the
data points are for eye guides. (b) Experimental and theoretical
values of transmission in the metamaterial. The theoretical curve
with $\gamma$ = 0.1$\omega_{sh}$ fits excellently with the
experimental data. (c) Phase velocities in the metamaterial. The
experimental data agree with the theoretical prediction from
equation (\ref{eq:sound-6}).} \label{fig:sh}
\end{figure}

Figure \ref{fig:sh}(a) shows acoustic propagation data in the
metamaterial for several frequencies. For the frequencies below 450
Hz (200 and 400 Hz) the sound intensity decayed exponentially with
the distance, $x$, from the boundary. Note that when the frequency
is lowered the curve approaches asymptotically to the theoretical
zero frequency curve. For the frequencies above 450 Hz (600, 800 and
1000 Hz) the sound waves propagated well with a slight decay due to
damping. The sound amplitudes are normalized to the value at $x$ =
0. The transmission data shown in figure \ref{fig:sh}(b) were
measured as the ratio of the pressure amplitudes at two different
positions; at $x$ = 0 and at $x$ = 1.3 m. Again it can be seen that
the sound waves above 450 Hz propagated well, but sounds below 450
Hz were completely blocked by the metamaterial. The theoretical
curves are from the calculation given below. The phase velocity in
the pass band was determined from the measured phase shifts with the
change of detector positions. Figure \ref{fig:sh}(c) shows the
experimental phase velocity together with the theoretical value
discussed below. The theory and the experimental results agree
excellently for both of the transmission and the phase velocity.

The novel behavior of the acoustic metamaterial is due to the motion
of air in the SH. The longitudinal wave motion in the tube is
affected by the motion of air moving in and out through the SH. The
air column in each hole has a mass given by $M = \rho \, l' \,
S$~\cite{aa}, where $\rho$ = 1.21 kg/m$^3$, $l'$ = 20 mm, and $S$ =
78.5 mm$^2$ are the density of air, the effective length, and the
area of the SH, respectively.

Now we present a theoretical model for the system. The air column
moves in and out with velocity $v$, driven by the pressure $p$ in
the tube according to the Newton's law, $p \, S = M \, dv/dt \ +\
b\,v$, where $b$ is the dissipation constant representing sum of the
drag loss and the radiation loss. As there are $n$ ($1/d$ = 14.2
m$^{-1}$) SHs per unit length, we can define the SH-mass density,
and SH-area-density as $\rho_{sh} \ = \ n \, M$, and $\sigma_{sh} \
= \ n \, S$, respectively. Because the spacing of the SH is much
smaller than the wavelength, the system can be regarded as a
homogenized medium. The SH then acts as a sink that modifies the
continuity equation in the tube:
\begin{equation} \label{eq:sound-3}
- \, \left( \frac{1}{B} \right) \frac{\partial {p}}{\partial t} =
{\bigtriangledown}\cdot \vec{u} \ + \ \left( \frac{\sigma_{sh}}{A}
\right) \, v,
\end{equation}
where $A$ is the cross section of the tube, and $\vec{u}$ is the
longitudinal velocity of the fluid inside the tube. Using the
harmonic expression $v = V \, e^{-i\,\omega\,t}$, this can be
simplified to
\begin{equation} \label{eq:sound-4}
{\bigtriangledown}\cdot \vec{u} = - \, \left( \frac{1}{B} -
\frac{\sigma_{sh}^2}{\rho_{sh} \, A \, \omega (\omega + i \gamma
)}\right)\ \frac{\partial {p}}{\partial t},
\end{equation}
where $\gamma$ is the damping term originated from the dissipation,
$b\,v$. The proportionality constant of the expansion
$({\bigtriangledown}\cdot \vec{u})$ to the pressure drop ${\partial
{p}}/{\partial t}$ is defined as the effective modulus,
\begin{equation} \label{eq:sound-5}
B_{eff}^{-1} = \left( \frac{1}{B} - \frac{\sigma_{sh}^2}{\rho_{sh}
\, A \, \omega (\omega + i \gamma )}\right)\ = B^{-1} \left( 1 -
\frac{\omega_{sh}^2}{\omega (\omega + i \gamma )}\right),
\end{equation}
where $\omega_{sh} = (B \, \sigma_{sh}^2 /A \, \rho_{sh})^{1/2}$.
The corresponding frequency, $f_{sh} (= \omega_{sh}/2 \, \pi)$ is
calculated to be 450 Hz from the parameters given above and the bulk
modulus of air $B = 1.42 \times 10^5$ Pa. The acoustic wave equation
obtained from equation (\ref{eq:sound-4}) and the Newton's equation,
$-{\bigtriangledown} p = \rho \, {\partial \vec{u}}/{\partial t}$,
gives the frequency dependent phase velocity,
\begin{equation} \label{eq:sound-6}
v_{ph} = \sqrt{\frac{B_{eff}}{\rho}} = \sqrt{\frac{B}{\rho \left(1 -
{\omega}_{sh}^2/\omega(\omega + i \gamma) \right)}} =
\frac{v_o}{\sqrt{1 - \omega_{sh}^2/\omega(\omega + i \gamma)}},
\end{equation}
where $v_o = \sqrt{B/\rho}$ is the speed of sound in air. Note that
equations (\ref{eq:sound-5}) and (\ref{eq:sound-6}) are formally
identical to equations (\ref{eq:sound-1}) and (\ref{eq:sound-2})
with the $B^{-1}$ and $\omega_{sh}$ corresponding to $\epsilon$ and
$\omega_{p}$ respectively. Equation (\ref{eq:sound-6}) predicts that
for the frequencies below $\omega_{sh}$, the acoustic waves do not
propagate but decays exponentially with distance, because the phase
velocities assume imaginary values. Above $\omega_{sh}$, the
acoustic waves propagate well. As the frequency is increased from
$\omega_{sh}$, the phase velocity decreases from a large value to
the asymptotic value, $v_o$. All these predictions are
experimentally verified as shown in figure \ref{fig:sh}(a) and (c).

From equation (\ref{eq:sound-6}), the dispersion $k(\omega)$ can be
calculated,
\begin{equation} \label{eq:sound-7}
k(\omega) = \frac{\omega}{v_{ph}} = \frac{\sqrt{\omega^2 -
\omega_{sh}^2/(1 + i \gamma / \omega)}}{v_o}.
\end{equation}
The imaginary component in the wave vector indicates decay of wave
amplitudes along the tube axis. The theoretical curves in figure
\ref{fig:sh}(b) show the intensity ratios calculated from equation
(\ref{eq:sound-7}) of the acoustic signals at two different
positions; at $x$ = 0 and at $x$ = 1.3 m.  The curve for the ideal
case of $\gamma$ = 0 shows pass/non-pass behavior for the acoustic
waves. The curve for $\gamma$ = 0.1$\, \omega_{sh}$ fits well the
frequency characteristics including the dissipation effect.

In summary, we presented fabrication of a non-local-resonator-type
of acoustic metamaterials. The theoretical model predicts negative
effective modulus in the frequency range from zero to a cut-off
frequency, $\omega_{sh}$. The experimental data agree excellently
with the theoretical predictions for the decay characteristics below
the cut-off frequency and the phase velocity in the pass band. We
expect the wide  spectral width of the present acoustic metamaterial
provide a basis for future research for acoustic double negative
metamaterials and further applications such as acoustic superlensing
and cloaking~\cite{19,20,21,22,23,24,25}.

\begin{acknowledgments}
The research was partially supported by The Korea Science and
Engineering Foundation (KOSEF R01-2006-000-10083-0).
\end{acknowledgments}

%\bibliographystyle{jasaauthyear}
%\bibliography{ad,stochastic,mm,dynamic}

\begin{thebibliography}{12345678}

\bibitem{1}  Veselago V G 1968 Sov. Phys. Uspekhi 10 509

\bibitem{2}  Pendry J B, Holden A J, Stewart W J and Youngs I 1996 Phys. Rev. Lett. 76 4773

\bibitem{3}  Pendry J B, Holden A J, Robbins D J and Stewart W J 1998 J. Phys.: Conden. Matter 10 4785

\bibitem{4}  Pendry J B, Holden A J, Robbins D J and Stewart W J 1999 IEEE Trans. Micr. Theory Tech. 47 2075

\bibitem{5} Smith D R, Padilla W J, Vier D C , Nemat-Nasser S C and Schultz S 2000 Phys. Rev. Lett. 84 4184

\bibitem{6} Fang  N, Xi D, Xu J, Ambati M, Srituravanich W, Sun C and Zhang X, 2006 Nature Mater. 5 452

\bibitem{7} Wang Z G \emph{et al}. 2008 J. Phys.: Condens. Matter 20 055209

\bibitem{8} Soukoulis C M, Zhou J, Koschny T, Kafesaki M and Economou E N, 2008 J. Phys.: Condens. Matter 20 304217

\bibitem{9} Cheng Y, Xu J Y and Liu X J 2008 Phys. Rev. B 77 045134

\bibitem{10} Hu X, Ho K -M, Chan C T and Zi J 2008 Phys. Rev. B 77 172301

\bibitem{11} Pendry J B 2000 Phys. Rev. Lett. 85 3966

\bibitem{12} Iyer A K and Eleftheriades G V 2008 Appl. Phys. Lett. 92 131105

\bibitem{13} Zhu J and Eleftheriades G V 2008 Phys. Rev. Lett. 101 013902

\bibitem{14} Schurig D \emph{et al}. 2006 Science 314 977

\bibitem{15} Chen H, Wu B -I, Zhang B and Kong J A 2007 Phys. Rev. Lett. 99 063903

\bibitem{16} Yang Z, Mei J, Yang M, Chan N H and Sheng P 2008 Phys. Rev. Lett. 101 204301

\bibitem{17} Li J and Chan C T 2004 Phys. Rev. E 70 055602(R)

\bibitem{18} Wood B and Pendry J B 2007 J. Phys.: Condens. Matter 19 07628

\bibitem{aa} Blackstock D T 2000 \emph{Fundamentals of Physical Acoustics} (New
York: Wiley)

\bibitem{19} Ambati M, Fang N, Sun C and Zhang X 2007 Phys. Rev. B 75 195447

\bibitem{20} Cummer S A and Schurig D 2007 New J. Phys. 9 45

\bibitem{21} Guenneau S, Movchan A, P\`etursson G and Ramakrishna S A 2007 New J.
Phys. 9 399

\bibitem{22} Cai L -W and S\`anchez-Dehesa J 2007 New J. Phys. 9 450

\bibitem{23} Chen H and Chan C T 2007 Appl. Phys. Lett. 91 183518

\bibitem{24} Cummer S A \emph{et al.} 2008 Phys. Rev. Lett. 100 024301

\bibitem{25} Torrent D and S\`anchez-Dehesa J 2008 New J. Phys. 10 063015


\end{thebibliography}

\end{document}